\documentclass[prb,floatfix,showpacs,amssymb,twocolumn]{revtex4}
\usepackage{graphicx}
\usepackage{dcolumn}
\usepackage{bm}

\newcommand{\bc}{\begin{center}}
\newcommand{\ec}{\end{center}}
\newcommand{\be}{\begin{equation}}
\newcommand{\ee}{\end{equation}}
\newcommand{\ba}{\begin{array}}
\newcommand{\ea}{\end{array}}
\newcommand{\beq}{\begin{eqnarray}}
\newcommand{\eeq}{\end{eqnarray}}

\begin{document}

\title{Strongly disordered Hubbard model in one dimension: \\
spin and orbital infinite randomness and Griffiths phases
}

\author{R. M\'elin}
\affiliation{
Centre de Recherches sur les Tr\`es Basses
Temp\'eratures (CRTBT), CNRS,
B. P. 166,
F-38042 Grenoble, France}

\author{F. Igl\'oi}
\affiliation{
Research Institute for Solid State Physics and Optics, 
H-1525 Budapest, P.O.Box 49, Hungary}
\affiliation{
Institute of Theoretical Physics,
Szeged University, H-6720 Szeged, Hungary}

\begin{abstract}
We study by the strong disorder renormalization group
(RG) method
the low-energy properties of the one-dimensional Hubbard model with
random-hopping
matrix-elements $t_{min}<t<t_{max}$, and with random on-site 
Coulomb repulsion
terms $0 \le U_{min}<U<U_{max}$.
There are two critical phases, corresponding to an infinite randomness
spin random singlet for strong interactions
($U_{min} > t_{max}$) and to an orbital infinite
randomness fixed point
for vanishing interactions ($U_{max}/t_{max} \to 0$).
To each critical infinite randomness 
fixed point is connected
a Griffiths phase,
the correlation length and dynamical exponent of which
have well defined asymptotic
dependences on the corresponding quantum control parameter. The theoretical
predictions for the scaling in the
vicinity of the critical points compare well to
numerical RG simulations.
\end{abstract}

\pacs{75.10.Nr, 75.50.Lk, 64.60.Ak}
\maketitle

\section{Introduction}

The Hubbard model is one of the simplest quantum many-body models of
interacting fermions.
Detailed exact information in one-dimension (1D) for the pure model
is provided by
Bethe Ansatz\cite{liebwu}, bosonization \cite{bosonization} and by factorizing
the wave-function\cite{ogatashiba}. For repulsive
interaction $U>0$, the 1D model is the simplest system with a Mott transition
at half-filling, having a finite charge gap $\Delta_c > 0$. At the same time
the
spin gap $\Delta_s$ is vanishing, and the spin-spin correlation function has
quasi-long-range order.

Quenched, {\it i.e.} time-independent disorder is an
unavoidable feature of real
materials and there are quasi-1D systems, such as tetracyanoquinodimethan
(TCNQ) compounds\cite{TCNQ} for which the Hubbard model with random
parameters is conjectured to be relevant.
Earlier theoretical studies \cite{Ma,sandvik,roemer}
based on the real-space
renormalization group (RG) method\cite{Ma}, 
quantum Monte Carlo (QMC) simulations\cite{sandvik}
and density matrix renormalization (DMRG)\cite{roemer}
are primarily interested in the effect of weak
random potentials
of strength $\epsilon$, in order to understand the metal-insulator transition
in the presence
of interactions. Repulsive interactions turn out to contrast
with attractive interactions:
like the non interacting model
\cite{localization},
the system with repulsive interactions is in
the insulating
phase\cite{metal_insulator} whereas a
metal-insulator transition is predicted 
for strong enough attractive interaction.
For repulsive interactions,
the charge gap is
reduced by the random potential and for strong enough $\epsilon$ 
there is a transition from a
Mott insulator (in which the charge gap is finite and the spin-spin 
correlation length is infinite)
to an Anderson insulator (with a vanishing charge gap and a
finite spin-spin correlation length).

Other investigations \cite{TCNQ_exp,FM}
of insulating disordered 1D systems, motivated
by the understanding of the NMP-TCNQ compound\cite{TCNQ_exp},
and more recently \cite{FM} by the inorganic spin-Peierls compound
CuGeO$_3$,
have relied on the evaluation of the uniform
paramagnetic susceptibility within strong disorder.
The proposed
effective model \cite{TCNQ} is a random exchange $S=1/2$
antiferromagnetic Heisenberg
spin chain. The RG introduced by 
Ma, Dasgupta and Hu\cite{mdh,review} to study this model was first implemented
numerically \cite{mdh}, and then analytically by Fisher \cite{fisherxx}.
According
to the RG results, the ground state
consists of a collection of randomly distributed singlets. Disorder
grows without limit as the energy is reduced. The resulting
singular behavior is typical of an
``infinite randomness'' fixed point:
the low-temperature uniform susceptibility diverges as\cite{fisherxx}:
\be
\chi(T) \sim \frac{1}{T |\ln T|^2}\;
\label{chi_RS}
\ee
which has been first found in numerical RG studies\cite{hirsch}. On the other
hand the experimentally
measured susceptibility in TCNQ materials is accurately described by
\cite{TCNQ_exp}
\be
\chi(T) \sim \frac{1}{T^{\beta}}\;,
\label{chi_exp}
\ee
with $\beta \approx 0.55 - 0.9$. The results of QMC simulations\cite{sandvik} are in qualitative
agreement with the power-law behavior in Eq. (\ref{chi_exp}).
A similar behavior was found more recently in the insulator o-TaS$_3$
where the power-law behavior in the susceptibility matches  accurately
\cite{TaS3} both
the power-law specific heat and the power-law 
electronic spin resonance spectroscopy signal \cite{Dumas}.

It is of importance to understand the low-energy properties of 
the Hubbard model in the presence of different types of randomness
and arbitrary interactions.
As shown in Ref.~\onlinecite{Artemenko}, dilute impurities
in an interacting quasi-1D conductor
can stabilize a ``bounded'' Luttinger liquid on ballistic finite size
segments in between two impurities that constitute infinite barriers
at low energy \cite{Kane-Fisher}.
We consider in the following a similar problem in the commensurate case
and in the regime of
strong disorder, likely to be relevant to a finite concentration
of impurities. The combined effects of
disorder in the hopping matrix-element and in the interaction
result in spin and orbital random singlets, and spin and orbital
Griffiths phases. The transitions and cross-overs between them
is the subject of our article.
For completeness we also consider
the possibility of diagonal disorder in the form of a random potential,
giving rise to an Anderson insulator.

The structure of the article is the following.
The random Hubbard model is
introduced in Sec. \ref{model}. The basic steps of the strong disorder
RG method together with the overall phase diagram of the model are presented in
Sec. \ref{SDRG}. Results of numerical renormalization group calculations are
given in Sec. \ref{results} and discussed in Sec. \ref{disc}.

\section{Model}
\label{model}
We start with the Hamiltonian of the 1D Hubbard model:
\begin{eqnarray}
{\cal H}&=&-\sum_{i=1}^L \sum_{\sigma=\uparrow,\downarrow} t_{i}
\left(c_{i,\sigma}^+ c_{i+1,\sigma} +
c_{i+1,\sigma}^+ c_{i,\sigma}\right)\nonumber \\ 
&+&\sum_i U_i n_{i,\uparrow} n_{i,\downarrow} +\sum_{i,\sigma} \epsilon_i c_{i,\sigma}^+
c_{i,\sigma}+{\cal H'}
\label{hamilton}
\end{eqnarray}
in which the extra term ${\cal H'}$
is generated during renormalization, so that ${\cal H}'=0$ in the
initial condition of the
physical model. 
The operators $c_{i,\sigma}^+$ and $c_{i,\sigma}$ create and annihilate
a spin-$\sigma$ fermion at site $i=1,2,\dots,L$,
and $n_{i,\sigma}=c_{i,\sigma}^+ c_{i,\sigma}$.
We restrict ourselves to repulsive on-site Coulomb interactions
being independent random numbers
such that $0\le U_{min} < U_i < U_{max}$, chosen in the distribution
\be
{\cal P}(U) = {\cal F}_{\alpha}(U,U_{min},U_{max})
\equiv \alpha \frac{(U_{max}-U_{min})^{-\alpha}}
{(U-U_{min})^{1-\alpha}}.
\label{power}
\ee
Similarly, the exchange integrals $t_{min}<t_i<t_{max}$
are independent and identically distributed
random numbers, and
we consider the distribution ${\cal P}(t) =
{\cal F}_{\alpha_t}(t,t_{min},t_{max})$.
We have also included in the Hamiltonian in Eq. (\ref{hamilton}) a
symmetrically distributed
random potential, $-\epsilon_{max} < \epsilon < \epsilon_{max}$.
The absolute value $|\epsilon|$ of the diagonal disorder potential
is drawn in the distribution 
${\cal P}(|\epsilon|) = {\cal F}_{\alpha_{\epsilon}}
(|\epsilon|,0,\epsilon_{max})$. 
We use the same exponent, $\alpha=\alpha_t=\alpha_{\epsilon}
=\alpha_U$ for the hopping, diagonal disorder potential
and interaction distributions. The parameter
$\alpha^{-2} =\textrm{var}[ \ln U]=
\textrm{var}[ \ln t]=\textrm{var}[ \ln |\epsilon|]$ measures
the strength of disorder for $U_{min}=t_{min}=0$, (${\rm var}[x]$ stands for the variance of $x$).
A uniform distribution corresponds to $\alpha=1$.
We restrict ourselves to the half-filled case:
$\sum_i \sum_{\sigma} n_{i,\sigma}=L$.

\section{Strong disorder RG, and overall phase diagram}
\label{SDRG}

To analyze the low-energy properties of the 1D
random Hubbard model we use the strong
disorder RG method\cite{review}, in which we first select
the interaction term (being the on-site Coulomb interaction, the hopping
integral or the on-site potential) with the
largest parameter in
the Hamiltonian in Eq. (\ref{hamilton}), that defines
the energy scale $\Omega$.
The largest coupling with strength
$\Omega$ is eliminated, which results in new parameters between
the remaining degrees of freedom, that are calculated
in perturbation.
We first set $\epsilon_{max}=0$ and
thus omit the random potential.

\subsection{Interaction dominated region}

\begin{figure*}
\includegraphics [width=1. \linewidth]{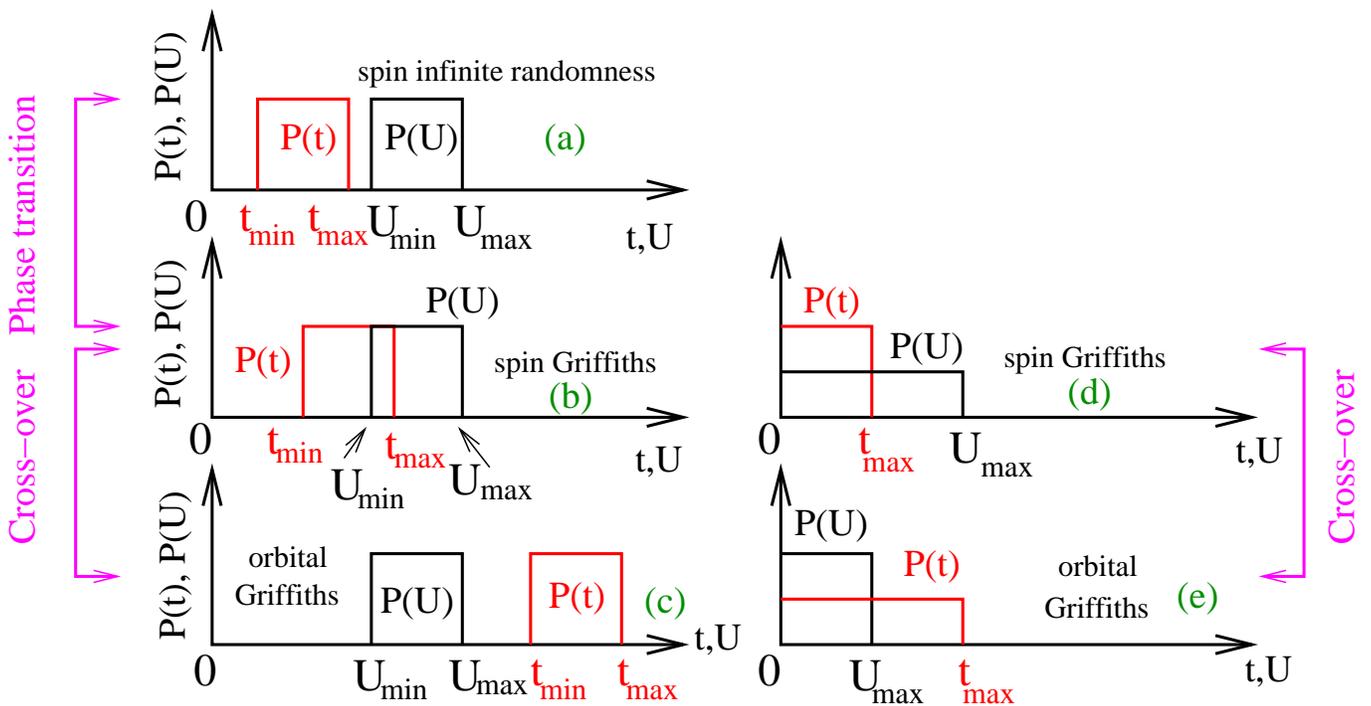}
\caption{(Color online) Schematic representation of the different hopping and
  on-site interaction energy distributions
  (drawn for $\alpha=1$), the corresponding phases and the
  transition or cross-over between them.
  \label{fig:distrib}
}
\end{figure*}

We start with a disorder in which the interaction plays the dominant role, 
so that
even the smallest interaction is larger than the possible maximal value
of the hopping
integral (see Fig.~\ref{fig:distrib}a):
\be
U_{min}>t_{max}\;.
\label{extreme}
\ee
In this case only interactions are decimated out in the first
stage of the RG, corresponding to $U$-transformations.

\subsubsection{$U$-transformation}

Let us consider the first decimation steps, when the largest
interaction term is, say,
$U_2=\Omega$, the hopping integral between sites $1$ and $2$
is denoted by $t_1$, and it is denoted by $t_2$ between sites
$2$ and $3$.
The double occupancy of site $2$ is forbidden (at least up to order of
$O(\max(t_1,t_2)/U_2)$) because of the large value of $U_2$.
The Hamiltonian is then projected on the subspace without
double occupancy. A virtual
exchange interaction is then
generated between the spin at site $2$ and between the neighboring
sites. If
site $1$ contains two fermions the exchange interaction is given in a
second-order perturbation calculation as:
\be
{\cal H'}_1 = \sum_{{\bf S}_1}\tilde{J}_1 {\bf S}_1.{\bf S}_2,
\quad \tilde{J}_1 \approx \frac{2t_1^2}{U_2}\;,
\label{tildej}
\ee
and similarly between sites $2$ and $3$.
New antiferromagnetic Heisenberg
exchange interactions are thus 
generated during renormalization. 
All the interactions are first decimated out
with the condition in Eq.  (\ref{extreme}).
If we perform the $U$ transformation at both
ends of a link we obtain for the final exchange term:
$\tilde{J}_1 \approx {2t_1^2}/{U_2}+{2t_1^2}/{U_1}$
and in ${\cal H'}_1$ in Eq. (\ref{tildej})
there is no sum over ${\bf S}_1$, since double occupancy
of site $1$ is also forbidden.
Thus after eliminating all the on-site Coulomb interactions we are left
with the Hamiltonian of the random antiferromagnetic Heisenberg chain:
$
{\cal H'}=\sum_{i=1}^{L} J_{i}
{\bf S}_i.{\bf S}_{i+1}$.
This model has been thoroughly studied by
strong disorder RG \cite{fisherxx}. The
corresponding RG-step is called as $J$-transformation.

\subsubsection{$J$-transformation}

In the decimation
procedure the two spins coupled by the strongest exchange coupling, say with $J_2=\Omega$, form
a singlet and are eliminated, while between the remaining sites, $1$ and $4$, a new exchange coupling is generated, given by
$
\tilde{J} \approx {J_1 J_3}/{2 J_2}\;,
$
where $J_1$ and $J_3$ are the original couplings joining to $J_2$.

\subsubsection{Random singlet spin infinite randomness phase}
\label{rsp}
The coupling distribution 
broadens without limit
as the renormalization goes on and the energy scale is lowered, 
corresponding to a spin
infinite randomness fixed point for the distributions
on Fig.~\ref{fig:distrib}a.
At this fixed point, the length-scale $L$ and the energy
scale $\Delta$ (the smallest gap) are related by
\be
-\ln \Delta \sim L^{1/2}\;,
\label{lnDelta}
\ee
and the distribution function
$P_L(\Delta)$ scales like
\be
\ln [L^{1/2} P_L(\Delta)] \simeq f(L^{-1/2} \ln \Delta)\;.
\label{inf_dis}
\ee
The ground state 
is the so called random singlet phase, consisting of effective
singlets between sites arbitrarily far apart.
The average spin-spin correlation function decays 
like $C(r) \sim r^{-2}$.
The low-temperature susceptibility has a Curie-like form with logarithmic corrections, as given
in Eq. (\ref{chi_RS}).
The predictions of the asymptotically exact RG have been confirmed by
numerical simulations\cite{henelius,ijr00},
see however Ref. [\onlinecite{stolze}].

\subsubsection{Random dimer phase}
\label{RDF}
To introduce the notion of Griffiths phase\cite{griffiths} that will be useful for
discussing the random Hubbard model,
we consider now an enforced dimerization, with
the couplings $J_o$ at odd positions, and $J_e$ at even
positions taken from different distributions.
The control parameter $\delta_{dim}$
is defined as
$
\delta_{dim}=[\ln J_o]_{\rm av} - [\ln J_e]_{\rm av}\;,
$
where $[ \dots ]_{\rm av}$ stands for
an average over quenched disorder. This
type of model
is obtained through renormalization of
the random Hubbard model if 
the hopping terms and/or the interactions in Eq. (\ref{hamilton})
are dimerized and if the relation in Eq. (\ref{extreme})
is satisfied.
The dimerized (non-random) Hubbard model is used to describe the
low-energy properties of
Bechgaard salts\cite{Bechgaard} and other quasi-1D systems\cite{quasi-1d}.

The dimerized random Heisenberg chain is renormalized to 
a zero energy fixed point\cite{fisherxx,HYBG96}, with short range spatial correlations 
{\it i.e.} the correlation length is $\xi < \infty$.
This is the so-called random dimer phase,
being a Griffiths phase consisting of
long singlet bonds involving the $J_o$ couplings and
short singlet bonds involving the $J_e$ couplings.
The average size $\xi$ of the latter 
is used to define the average correlation length, which
behaves for small dimerization as
\be
\xi \sim \delta_{dim}^{-2}\;.
\label{xi_dim}
\ee
On the other hand, the length of the long singlet bonds is proportional to the
size $L$ of the system,
 and the low-energy excitations goes to zero as
\be
\Delta \sim L^{-z}\;.
\label{z}
\ee
Here $z<\infty$ is the dynamical exponent which depends
on the value of $\delta_{dim}$,
thus on the distribution of disorder. 
Close to the random singlet phase, $|\delta_{dim}| \ll 1$,
the dynamical exponent diverges as
\be
z \sim \delta_{dim}^{-1}\;.
\label{z_dim}
\ee
If the low-energy excitations are localized, which is the case
in the random dimer phase,
the gap distribution $P_L(\Delta)$ scales like
\be
P_L(\Delta)=L^z g(L^z \Delta) \sim \Delta^{-1+1/z}\;
\label{P_Gr}
.
\ee
The low-temperature uniform susceptibility in the random dimer phase 
behaves like
$
\chi(T) \sim {1}/{T^{1-1/z}}\;,
$
which is in the same form as the experimental result in Eq. (\ref{chi_exp}), with $\beta=1-1/z$.
In the random singlet phase, Eqs.(\ref{z})
and (\ref{P_Gr}) are replaced by Eqs.(\ref{lnDelta}) and (\ref{inf_dis})
respectively.

\subsection{Hopping dominated region}
\label{hphase}

Some hopping matrix elements are
also decimated as
$\Omega$ is lowered below $t_{max}$,
if the condition set by Eq. (\ref{extreme})
on the strength of the on-site Coulomb interaction 
is not satisfied, which we consider now.

\subsubsection{$t$-transformation}

If the energy $\Omega$ corresponds to a hopping term, say $\Omega=t_2$,
then two fermions (one with $\sigma=\uparrow$ and 
the other with $\sigma=\downarrow$) are localized on the
bond between sites $2$ and $3$, and these sites are eliminated. 
A small hopping term is generated
between
the remaining sites $1$ and $4$, which is given in second order
perturbation by
$
\tilde{t} \approx -{t_1 t_3}/{t_2}\;.
$
This transformation has an effect on the value of
the interaction term, say at site $1$, provided $U_1$ has not yet been decimated out.
For its renormalized value we obtain:
\be
\tilde{U}_1 \approx U_1 -\frac{t_1 t_3}{t_2}\;.
\label{tildeU}
\ee
Similarly the random potential at site $1$ is modified and transformed as
\be
\tilde{\epsilon}_1 \approx \epsilon_1 + \frac{t_1 t_3}{t_2}\;.
\label{tildeeps}
\ee
Formula analogous to Eqs.(\ref{tildeU}) and (\ref{tildeeps})
hold also at site $4$.

Finally, if two $J$-couplings (denoted
by $J_1$ and $J_3$) have already been
generated at sites $1$ and $4$,
an effective exchange interaction is also generated between sites
$1$ and $4$:
\be
\tilde{J} \approx \frac{J_1 J_3}{16 |t_2|}\;.
\label{tildejt}
\ee
However, if $J_1=0$ and/or $J_3=0$ we obtain to third order:
\begin{equation}
\tilde{J} = 12 \frac{|t_{1}|^2 |t_{3}|^2}
{|t_{2}|^3}\;.
\label{tildej3}
\end{equation}
The $t$-transformation has an influence on the renormalized
value of the other parameters, in particular on the on-site Coulomb interaction.
In particular, both $t$- and $U$-transformations
are generated at some stage of the renormalization for
any finite value of $U_{max}>0$. 

\subsubsection{Orbital
infinite randomness fixed point: $U_{max}=\epsilon_{max}=0$.}
\label{OIDFP}
The problem reduces to the random
tight-binding model and the RG process involves 
solely $t$-transformations
if $U_{max}=0$ (and $\epsilon_{max}=0$).
The fixed point of the RG
is now an orbital infinite randomness fixed point\cite{mdh02,mdi05}, which is
isomorph to the fixed point of the random
antiferromagnetic Heisenberg chain, as discussed in Sec. \ref{rsp}. In
particular the relations
in Eqs.(\ref{lnDelta}) and (\ref{inf_dis}) remain valid.
The ground state of the
system is
made of $t$-frozen pairs of fermions, where the length of a pair can be
arbitrarily large.

\subsection{Interplay between interaction and hopping}

In the general situation $U_{min}<t_{max}$, $U_{max} > 0$ where the $t$-
and $U$-distributions overlap (this corresponds to Figs.\ref{fig:distrib} b,c,d and e),
both $U$ and $t$ terms are decimated
during renormalization
and $J$-transformations are also carried out.
The ground state of the system is a mixture of $J$-singlets
and frozen $t$-pairs and the average size $\xi_J$ of the singlets,
and that $\xi_t$ of the
$t$-pairs are finite. The largest of the two defines the
correlation length. At the same time 
the energy scale decreases to zero with the size of the
system as in Eq. (\ref{z}).
We are thus in a Griffiths phase, analogous
to the random dimer phase described in Sec. \ref{RDF}, which
can be divided
in two regions.
The on-site Coulomb interaction plays the dominant role
in the ``spin Griffiths phase'' corresponding to $\xi_J>\xi_t$
for the coupling distributions on Fig.~\ref{fig:distrib}b and d,
whereas hopping is dominant in the ``orbital Griffiths phase'',
corresponding to $\xi_t>\xi_J$ for the coupling distributions
on Fig.~\ref{fig:distrib}c and e.
There is a cross-over, but no sharp transition
in between these two Griffiths phases, similarly
to that observed in other random quantum models,
such as the random dimerized $S=1$
chain\cite{S1} or random Heisenberg ladders\cite{ladder}.
In the following subsections we analyze
the properties of the Griffiths phases in the vicinity of the infinite
randomness fixed
points. Numerical results far from the
random critical points are presented afterwards.

\subsubsection{Spin Griffiths phase: $0<t_{max}-U_{min} \ll U_{max}$}
\label{SGP}
Let us start to analyze the properties of the system in the vicinity of the
random singlet fixed point, when $0<t_{max}-U_{min} \ll U_{max}$.
We define the control-parameter $\delta_U$ as the
fraction of non-decimated $U$-terms at $\Omega=t_{max}$:
\be
\delta_U=\int_{U_{min}}^{t_{max}} P(U) {\rm d} U =
\left( \frac{t_{max}-U_{min}}{U_{max}-U_{min}}\right)^{\alpha}\;.
\label{delta_U}
\ee
Here the second relation holds for the power-law distribution in Eq. (\ref{power}).

As we discussed in Sec. \ref{hphase},
some $t$-terms
are also decimated out 
as $\Omega$ is lowered below $t_{max}$,
and the density $\rho_t$ of the frozen $t$-terms is
given by
$\rho_t \sim \delta_U$. The typical length-scale in the system is given by the
typical distance
between two frozen $t$-terms, and is thus
\be
\xi_{typ} \sim 1/\rho_t \sim \delta_U^{-1}\;.
\label{xi_typ}
\ee
On the other hand the average correlation length is given by the average
distance between spins
forming a singlet through a $J$-coupling. In the random singlet phase,
{\it i.e.} for $\delta_U=0$, the singlets are formed either from
odd or from even bonds. For $\delta>0$ the correlated segments are broken
if two neighboring $t$-terms are frozen, which leads to a change of the
parity of the singlets, see the discussion in Sec. \ref{RDF}.
Thus we obtain for the average correlation length
\be
\xi \sim 1/\rho_t^2 \sim \delta_U^{-2}\;.
\label{xi_U}
\ee
Finally, one considers the typical value of the gap, which is related to the value of the gap
in the random singlet phase of size $\sim \xi$. Eqs.
(\ref{lnDelta}), (\ref{z}) and (\ref{xi_U}) lead to
\be
z \sim \delta_U^{-1}\;.
\label{z_U}
\ee
Now comparing Eqs.(\ref{xi_U}) and (\ref{z_U}) with those of
Eqs.(\ref{xi_dim}) and (\ref{z_dim})
in the random dimer phase we
conclude that $\delta_U$ in Eq. (\ref{delta_U}) plays the
role of the control parameter in the spin Griffiths phase.

\subsubsection{Orbital Griffiths phase - $0<U_{max} \ll t_{max}$}
\label{OGP}
Second, we consider the behavior of the system in the vicinity of the orbital infinite
randomness fixed point, $0<U_{max} \ll t_{max}$. In the initial steps of the renormalization
solely $t$-terms are decimated out, until the energy-scale is lowered below $\sim U_{max}$,
when also $U$-terms are eliminated. The density of decimated $U$ sites is
given by $\rho_U \sim |\ln U_{max}|^{-2}$, which
follows from Eq. (\ref{lnDelta}), with the
correspondences $\Delta \sim U_{max}$ and
$L \sim \rho_U^{-1}$. We have thus for the correlation length $\xi_t$
(the average size of the $t$-singlets):
$\xi_t \sim \rho_U^{-1} \sim |\ln U_{max}|^{2}$.
Comparing to Eq. (\ref{xi_dim}) in the random dimer
phase we identify the control parameter in the orbital Griffiths phase as:
\be
\delta_t = |\ln( U_{max}/t_{max})|^{-1}\;.
\label{delta_t}
\ee
Repeating the arguments used in the vicinity of the random singlet phase we
obtain that the
relations in Eqs.(\ref{xi_typ}), (\ref{xi_U}) and (\ref{z_U}) 
remain valid by
simply replacing $\delta_U$
by $\delta_t$.

\subsection{Role of the random potential - Anderson vs. Mott insulator phases}

In order to determine
the overall phase diagram of the random Hubbard model we
consider also the role of the random potential $\{\epsilon_i\}$ which
plays a special role in the renormalization in 1D. If at
some stage of the renormalization one random potential term is decimated, say
$\epsilon_2=\Omega$, then
two fermions are frozen at site $2$, and this site is eliminated. This transformation does not influence
the value of the parameters at the neighboring sites
because no hopping is generated between the remaining sites
$1$ and $3$. Transport is thus blocked at this site.
The
random potentials are just slightly modified by 
other $t$-transformations (see in Eq. (\ref{tildeeps})).
The random potentials are thus also decimated at some stage of the RG
if $\epsilon_{max} >U_{max} \ge 0$. 
As a consequence a finite fraction of sites is
frozen and the system scales into the Anderson insulator phase.
In the other limiting case, $\epsilon_{\max} < U_{min}$, the behavior of the system is dominated
by the on-site Coulomb repulsion, thus no fermions are frozen due to the
random potential and the system scales into the Mott insulator phase.

\subsection{Phase diagram}

\begin{figure}
\includegraphics [width=.9 \linewidth]{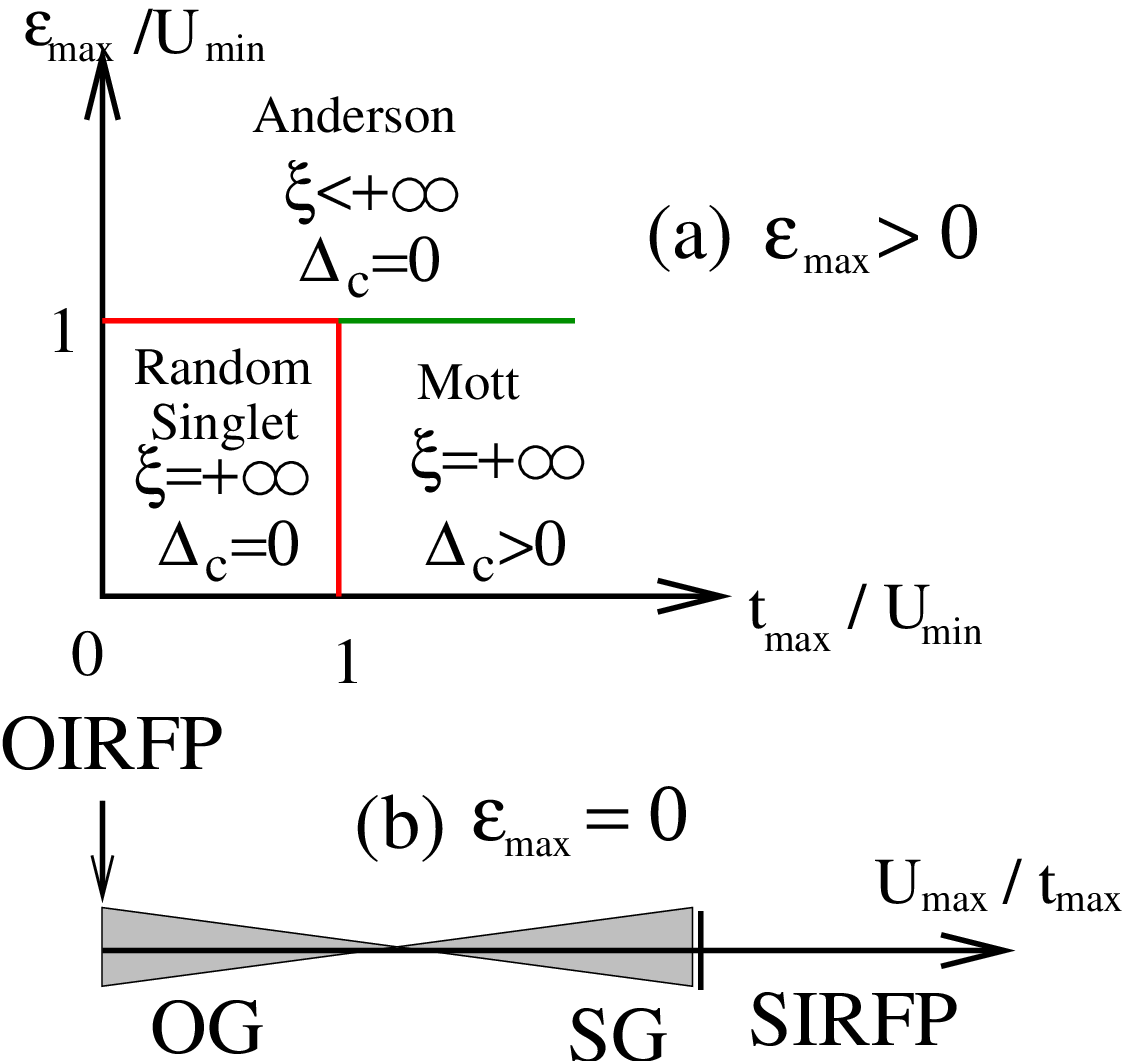}
\caption{(Color online) Schematic phase diagram of the random Hubbard chain. a) Including a
  random
  potential and with weak disorder in $U$ and $t$. b) Without random potential
  and with
  strong disorder in $U$ and $t$. This latter corresponds to the numerical work
  in the paper. OIRFP
  stands for orbital
  infinite randomness, OG for orbital Griffiths (for the distributions on
  Fig.~\ref{fig:distrib}c), SG for spin Griffiths
  (for the distributions on Fig.~\ref{fig:distrib}b), SID for
  spin infinite randomness
  (for the distributions on Fig.~\ref{fig:distrib}a).
  See the text for the properties of the different phases.
  We suppose as in Fig.~\ref{fig:distrib}
  coupling distributions
  with constant values of $U_{max}-U_{min}$ and
  $t_{max}-t_{min}$ while $U_{max}$ and $t_{max}$ are varying.
  \label{fig:diag1}
}
\end{figure}

We conclude
this section by presenting the phase diagram of the random Hubbard model
in 1D, which constitutes the main result of this paper.
First
we consider in Fig.~\ref {fig:diag1}a
the effect of a random potential, when the disorder
in the other terms is weak, so that $(t_{max}-t_{min})/(t_{max}+t_{min})
\sim (U_{max}-U_{min})/(U_{max}+U_{min})=D_{t,U} \ll \epsilon_{max}$. Previous
numerical
work\cite{sandvik,roemer} considered non-random $U$ and $t$, thus
$D_{t,U}=0$. Here we extend these results to
weak disorder.
The random potential plays a dominant role for $\epsilon_{max}/U_{min}>1$,
when the system is in the Anderson insulator phase, with a finite
spin-spin correlation length
$\xi<\infty$ and with a vanishing charge gap $\Delta_c=0$.
If $D_{t,U}=0$ and $\epsilon_{max}/U_{min}<1$ the system is
in the Mott insulator phase, $\xi=\infty$ but with $\Delta_c>0$. We expect
this scenario to remain
valid even for weak disorder ($D_{t,U} \ll \epsilon_{max}$) provided
hopping dominates
over Coulomb repulsion, thus $t_{max}/U_{min}>1$. If, however the
on-site
Coulomb repulsion is sufficiently strong ($\epsilon_{max}/U_{min}<1$ and
$t_{max}/U_{min}<1$)
the system is in the spin random singlet phase, see Sec. \ref{rsp}, with a divergent spin-spin correlation length, $\xi=\infty$ and with a vanishing charge gap, $\Delta_c=0$. Transition between the Mott
and the Anderson phases is found to be controlled by a conventional random
fixed point\cite{sandvik,roemer},
whereas transitions from the spin random singlet phase are likely of
infinite randomness type.

In Fig.\ref {fig:diag1}b we present the phase diagram without a random potential, $\epsilon_{max}=0$,
but for sufficiently strong disorder in $U$ and $t$ as a function of $r=U_{max}/t_{max}$.
For $U_{max}/t_{max}=0$ the system is in the orbital infinite randomness fixed point, the
properties of which are described in Sec. \ref{OIDFP}. In the other limiting case with
$U_{min} > t_{max}$, which corresponds to
 $U_{max}/t_{max} > r_c>0$, the system is in
the spin random singlet phase (see in Sec. \ref{rsp}). For $0<U_{max}/t_{max} < r_c$ the
system is in the Griffiths phase, which is divided
in an orbital Griffiths phase (Sec. \ref{OGP}) for
$0<U_{max}/t_{max} \ll r_c$, and in
a spin Griffiths phase (Sec. \ref{SGP}) for
$0 \ll U_{max}/t_{max} < r_c$. In the following section we study now
numerically the properties
of the phase diagram in Fig.\ref {fig:diag1}b.

\section{Numerical results}
\label{results}

We consider 
typically $250000$ independent random samples taken from the distribution in
Eq. (\ref{power})
to evaluate the dynamical exponent
in the numerical implementation of the strong disorder RG method.
We use a smaller statistics of $\sim 10000$ samples for the
evaluation of the correlation lengths.
The numerical results are presented only for the uniform distribution
with $\alpha=1$, but we obtained similar results for smaller values of
$\alpha$.
The length of the chain is varied up to $L=2048$ and
the decimation is performed up to the last remaining particle or spin
singlet. The gap $\Delta$ at the last step
of renormalization
is identified as the gap of the random chain and we have also evaluated
the length scales $\xi_t$ and $\xi_J$ from the density of the decimated
$t$ and $J$ terms respectively.

\subsection{Qualitative features of the RG flow}

To illustrate the behavior of the RG flow we have calculated the fraction
$n_t(\Gamma)\delta\Gamma$ of $t$-transformations,
the fraction $n_U(\Gamma)\delta\Gamma$ of $U$-transformations, 
and the
fraction $n_J(\Gamma)\delta\Gamma$
of $J$-transformations in an interval $[\Gamma,\Gamma+\delta \Gamma]$,
where $\Gamma=-\ln \Omega$ is the log-energy scale. 
The quantities are normalized in such a way as
$n_t(\Gamma)+n_U(\Gamma)+n_J(\Gamma)=1$.

\begin{figure}
\includegraphics [width=.8 \linewidth]{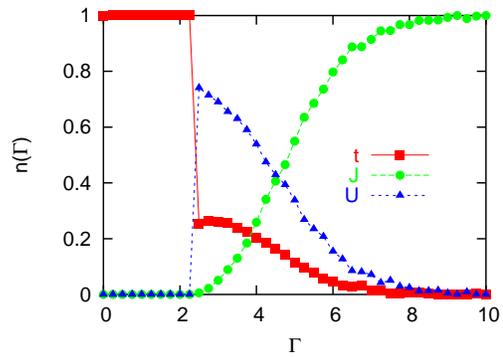}
\caption{(Color online) The fractions of different RG transformations as a function of
  the log-energy scale, $\Gamma=-\ln \Omega$, for the random 1D chain. 
  \label{fig:Fig4a}
}
\end{figure}

The variations of $n_t(\Gamma)$, $n_U(\Gamma)$, $n_{J}(\Gamma)$
for the 1D chain are shown on Figs.~\ref{fig:Fig4a}
for $U_{max}=0.1$, $t_{max}=1$ and $\alpha=1$
with $U_{min}=t_{min}=0$. 
As seen in this figure only $t$-terms are decimated out
at the beginning of the RG, and
the decimation of the $U$-terms starts only when $\Omega$
is lowered below $U_{max}$. 
$J$-terms are also generated through $U$-decimations,
which are then also decimated.
The $U$- and $t$-terms gradually die out
by further decreasing
$\Omega$ and one is left with the $J$-transformations at low energy.
 
\subsection{Properties of the Griffiths phases}

\subsubsection{Dynamical exponent}

\begin{figure}
\includegraphics [width=.8 \linewidth]{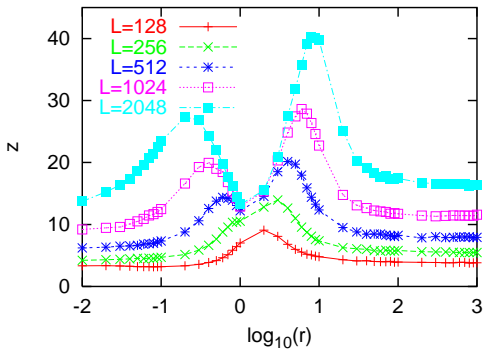}
\caption{(Color online) Dynamical exponent as a function of $r=U_{max}/t_{max}$ for different
  system sizes ($U_{min}=t_{min}=0$, $\alpha=1$).
\label{fig:Fig4bis}
}
\end{figure}

The dynamical exponent $z$ is
determined from the distribution of the gaps,
which in
the Griffiths phases follows the scaling form given in Eq. (\ref{P_Gr}). For a
finite system
this procedure results in effective $L$-dependent dynamical exponents, that
are then extrapolated to $L=\infty$.
The calculated effective dynamical exponents for the
power-low
distribution in Eq. (\ref{power}) with $t_{min}=U_{min}=0$ and $\alpha=1$ for
various
values of $r=U_{max}/t_{max}$ are shown in Fig.~\ref{fig:Fig4bis}.
For these distributions
the infinite randomness fixed points are located at $r=0$
(orbital infinite randomness
fixed point), and at $r=\infty$ (spin random singlet,
infinite randomness phase). The dynamical exponent $z$
is formally infinite at both fixed points,
which is compatible with the fact that the effective
$z$ is increasing with $L$ with no sign of saturation.
It is also interesting to note that the maxima of the effective
exponent in the orbital Griffiths phase
for a size $L$ are approximately the same as the
maxima of the same curve in the spin Griffiths phase, however with a size
$L/2$. This is because the spin Griffiths phase involves
approximately the double of RG steps
(in the spin Griffiths phase one makes also a set of extra
$U$-transformations).

\begin{figure}
\includegraphics [width=.8 \linewidth]{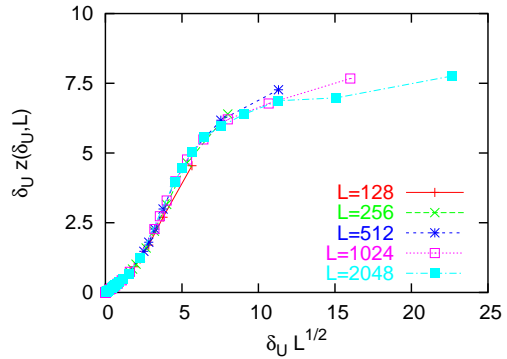}
\caption{(Color online) Scaling plot of the dynamical exponent in Fig.\ref{fig:Fig4bis} in the
spin Griffiths phase, with $\delta_U$ in Eq. (\ref{delta_U}).
\label{fig:Figsup1}
}
\end{figure}

\begin{figure}
\includegraphics [width=.8 \linewidth]{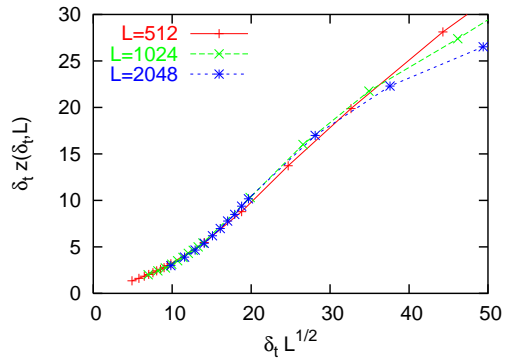}
\caption{(Color online) The same as in Fig.\ref{fig:Figsup1} in the
orbital Griffiths phase, with $\delta_t$ in Eq. (\ref{delta_t}).
\label{fig:Figsup2}
}
\end{figure}

To have a qualitative analysis of the data we recall that close to the infinite
randomness fixed points the effective exponent
$z(\delta,L)$ obeys the scaling
form:
\be
z(\delta,L)=\delta^{-1} \tilde{z}(L^{1/2} \delta)\;,
\label{z_scal}
\ee
in which we have incorporated relations in Eq.  (\ref{z_dim}) and in
Eq. (\ref{xi_dim}).
Furthermore, the scaling function behaves like
$\tilde{z}(x) \sim x$ for small $x$,
leading to an $\delta$-independent $\sim L^{1/2}$ effective dynamical exponent
at the critical points.
The scaling form in Eq. (\ref{z_scal}) is well verified
in the vicinity of the two
infinite
randomness fixed points, keeping in mind that the control parameter $\delta$
is given
in Eqs.(\ref{delta_U}) and (\ref{delta_t}) for the spin and the orbital
Griffiths phases
respectively. We obtain a satisfactory
agreement between the
numerical results and the analytical calculations in
both regions (see Figs.~\ref{fig:Figsup1} and~\ref{fig:Figsup2}).

\subsubsection{Correlation length}

\begin{figure}
\includegraphics [width=.9 \linewidth]{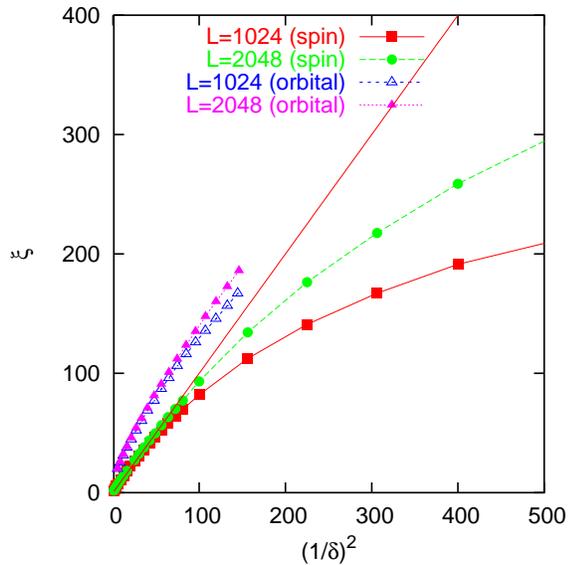}
\caption{(Color online) 
Correlation length in the spin Griffiths phase
and the orbital Griffiths phase 
given by the average
length of the spin and orbital singlets respectively
as a function the inverse of the square of the control parameter
(see Eqs. (\ref{delta_U}) and
(\ref{delta_t}) respectively).
The theoretical predictions correspond to the solid line
$\xi=\delta^{-2}$.
\label{fig:xi}
}
\end{figure}

The correlation length is finite in the Griffiths phases and given
either by the average length $\xi_J$ of the spin singlets in the spin
Griffiths phase, or by that
$\xi_t$ of the orbital singlets in the orbital
Griffiths phase. In terms of the appropriate control parameter, given either
in Eq. (\ref{delta_U}) or in Eq. (\ref{delta_t}), the correlation lengths are
divergent in the vicinity of the infinite randomness fixed points, as given
in Eq. (\ref{xi_dim}). This relation has been 
verified by numerical simulations (see
Fig.\ref{fig:xi}),
in which -- in order to get rid of finite size effects -- we used large finite
systems with length up to $L=2048$. For the spin Griffiths-phase 
we used a distribution of the form in Fig.\ref{fig:distrib}b, whereas for the
orbital Griffiths phase the distribution is given
in Fig.\ref{fig:distrib}e.
As seen in the figure satisfactory agreement with the theoretical prediction
is found, at least for small enough $\delta$ and for large $L$.

\section{Discussion and conclusion}
\label{disc}
In this paper we have studied the low-energy, low-temperature properties of the
one-dimensional Hubbard model in the presence of quenched disorder. 
Contrary to
previous studies\cite{Ma,sandvik,roemer} in which disorder was
comparatively weak
and realized in the form of a random potential, we considered here 
disorder both
in the hopping integrals and in the on-site Coulomb repulsion. Due to
the sufficiently
strong disorder, the system is always gapless, which means
that the
low-energy properties are controlled by zero-energy fixed
points, for
any value of the quantum control parameter.
In the framework of the strong
disorder
renormalization group that we have adapted here to the random Hubbard
model, we have
identified two distinct critical phases, which are controlled by infinite
randomness fixed
points. For dominant on-site repulsion the system is a spin random singlet
infinite randomness
phase, whereas
for vanishing on-site
Coulomb interaction the system corresponds to a random tight-binding model
the properties of which are controlled by an orbital infinite randomness fixed point.

The two infinite randomness
critical phases are separated by two different Griffiths phases,
between
which there is a smooth cross-over, but no sharp transition. 
The strong disorder RG method is expected to provide
asymptotically exact results
close to the critical phases, where
the correlation length is divergent.
We have made analytical predictions, both
for the
divergence of the correlation length and that of the dynamical exponent. These
results were
compared to large scale numerical RG calculations and satisfactory
agreement is found.

To conclude, we mention possible extensions of our work. It is of
interest to
study the weak-to-strong disorder effects in the random Hubbard chain. For
example the
charge gap in the Mott insulator phase is expected to be robust against weak
disorder, but
strong disorder will destroy it, as shown by our RG results. 
It would be also of interest
to study the combined effects of disorder realized at the same time in
different
parameters (potential, hopping and interaction).
Finally, one may consider
extensions of the model
to higher dimensions. A new physics is expected already in two dimensions
because the interaction dominated phase that maps to a
random Heisenberg
model has a conventional random fixed point\cite{2d}, whereas for vanishing
on-site Coulomb repulsion
the system is in a logarithmically infinite randomness fixed
point\cite{2dtight}. However, for a quasi-one dimensional system,
made of weakly coupled chains, the transverse exchange couplings
of the low energy Heisenberg model are much smaller
than the longitudinal ones because of Eq. ~\ref{tildej}.
The one-dimensional spin Griffiths phase due
to strong on-site Coulomb
interactions is thus expected also in the quasi-one-dimensional
system with sufficiently small interchain couplings.

\section*{Acknowledgments}
F.I. acknowledges useful discussions with K. Penc and A. Sandvik.
This work has been supported by the French-Hungarian cooperation
programme Balaton (Minist\'ere des Affaires Etrang\`eres - OM), the
Hungarian National Research Fund under grant No OTKA TO37323, TO48721, K62588, MO45596 and M36803.
The Centre de Recherches sur les Tr\`es Basses Temp\'eratures is
associated with the Universit\'e Joseph Fourier.

\end{document}